# Visualizing individual microtubules using bright-field microscopy

Braulio Gutiérrez-Medina[a), b)]
*Department of Biology, Stanford University, Stanford, CA 94305*
Steven M. Block[c)]
*Department of Biology, Stanford University, Stanford, CA 94305 and*
*Department of Applied Physics, Stanford University, Stanford, CA 94305*

Microtubules are filament-shaped, polymeric proteins (~25 nm in diameter) involved in cellular structure and organization. We demonstrate the imaging of individual microtubules using a conventional bright-field microscope, without any additional phase or polarization optics. Light scattered by microtubules is discriminated through extensive use of digital image-processing, thus removing background, reducing noise and enhancing contrast. The setup builds on a commercial microscope, with the inclusion of a minimal and inexpensive set of components, suitable for implementation in the student laboratory. We show how this technique can be applied to a demonstrative biophysical assay, by tracking the motions of microtubules driven by the motor protein kinesin.

## I. INTRODUCTION

Since its conception, the light microscope has been a powerful tool in biology, providing means to study a multitude of essential processes occurring within the cell. Although resolution in light microscopy is limited to ~$\lambda/2$ due to diffraction effects[1], where $\lambda$ is the wavelength of the illumination light, cellular macromolecules (which measure less than 50 nm in size) can be routinely visualized by a number of techniques that involve fluorescence, phase or polarization effects. One cellular component whose study has greatly benefited from advances in light microscopy is the microtubule[2]. Polymerized from the protein tubulin, microtubules (MTs) are long, hollow filaments involved in dictating cellular structure, organization and division[3]. The dimensions of these biopolymers are ~25 nm in diameter, corresponding to only 1/20 of the wavelength of visible light, and several microns in length. Specialized light microscopy methods that visualize individual MTs include: fluorescence[4], dark field[5], phase contrast[5], and polarization[6] microscopy, and Nomarski-differential interference contrast (DIC) imaging[7]. The quality of images obtained by any of these means is often improved by subsequent video-processing, where analog or digital manipulation of video frames subtracts background, reduces noise and increases contrast[8]. Using video-enhanced, Nomarski-DIC imaging, it is even possible to visualize the flagella of bacteria (~15 nm diameter)[9]. While these microscopy methods are widely used in biophysical research, they are often prohibitively expensive for implementation in the classroom.

Here, we show that detection of scattered light with the bright-field microscope is enough to offer a simple alternative for microtubule localization. We take advantage of the fact that the intensity of scattered light due to a small, elongated cylinder scales as: $I \propto (a/\lambda)^4$, where $a$ is the radius of the cylinder[10]. This contrasts with the $I \propto r^6$ dependence for a sphere of radius $r$, making MTs more accessible for bright-field imaging over spherical nanoparticles (such as quantum dots). To visualize individual, unstained microtubules we image light scattered by MTs onto a CCD camera, and images are further enhanced using computer software.

We developed our setup around a commercial microscope, with simple and inexpensive additions that contribute to robust imaging: a light emitting diode (LED) for illumination, a student-available microscope objective, a board-level CCD camera, and freeware to perform subsequent image manipulation. The experiments proposed in this paper will allow students to become acquainted with key elements of light scattering, the importance of proper alignment of the microscope, and the advantages of image processing to achieve visualization of an essential, sub-diffraction cellular component. Furthermore, we provide detailed instructions on how to prepare biological samples to perform an *in vitro* biophysical experiment, by using the bright-field microscope to quantitatively measure MT displacement when driven by the molecular motor kinesin.

## II. EXPERIMENTAL SETUP

Bright-field microscopy is the simplest configuration for specimen illumination of the light microscope, where illumination light travels through the condenser, reaches the sample, and both scattered and unscattered light are collected by the objective and subsequently imaged onto the eyepiece or a camera (Figure 1a). We use this configuration throughout this paper, without any additional optical components such as filters, polarizers or prisms. We describe below each of the elements involved in our setup.

**A. Microscope**

In order to optimize image quality and contrast, the microscope must allow independent adjustment of various elements: the illumination system (light source and collector lens position, and field iris opening), the condenser (condenser lens position and condenser iris opening), and the objective position. To satisfy these requirements, we used a commercial, inverted microscope (Axiovert 100, Zeiss), featuring a video camera port and a home-made XY translation stage to support the sample holder (Figure 1b). Although a commercial inverted microscope is expensive (> $10,000), we anticipate that students can have access to an equivalent instrument, or be able to build the necessary setup using basic optical and mechanical components[11-13]. The inverted configuration is not required, but it provides ample space for further addition of optical elements to implement other techniques used in biophysical studies, such as laser trapping or fluorescence microscopy[13].

**B. Illumination**

In the last few years, light-emitting diodes (LEDs) have emerged as an inexpensive, high-quality alternative to discharge lamps for light microscopy illumination[14, 15]. LEDs have recently been used to visualize individual MTs, using Nomarski-DIC optics, and it was found that short illumination wavelengths provide better images[16], consistent with the wavelength-dependence of scattered intensity by microscopic filaments. For these reasons, we used a blue LED (peak emission at $\lambda =$ 455 nm) for illumination, manufactured by Luxeon (V Star, Royal Blue Lambertian, Cat. #: LXHL-LR5C, $25). This is a high-power LED array (up to 700 mW of emitted power), designed by the manufacturer to have a radiation profile proportional to the cosine of the forward angle, thus making it possible to approximate the emitter as a point source. The array was driven at constant current (~650 mA) using a simple custom-made electronic servo loop[17] (Figure 2). Similar LED control circuits have been published[14]. The LED was mounted on the cap of a 2″ lens tube (Thorlabs), and attached to a heat sink. Inside the tube, a 2″-diameter lens of $f$ = 60 mm focal length collimated the emitted light. The lens tube ensured the LED to be centered with respect to the collimating lens, whereas internal threads allowed adjustment of the focusing depth.

**C. Objective and condenser**

As mentioned previously, single microtubules can be visualized using phase, fluorescence or polarization effects —methods that require the use of microscope objectives with minimal chromatic aberrations and reduced stress that avoid the introduction of image and polarization distortions. The extreme care needed in the production of such objectives makes them very expensive (> $10,000), not to mention the polarizers and prisms often required for the control and detection of emitted or scattered light. In contrast, our technique relies purely on distinguishing scattered from transmitted

illumination light, which we do using digital processing. It is therefore possible to use an inexpensive, 'student-grade' objective. We tested two different oil-immersion objectives from Edmund Optics: the 100X DIN Plan Commercial Grade (N.A. = 1.25, Cat. #: NT43-909, $199), and the 100X DIN Semi-Plan, Spring Loaded International Standard (N.A. = 1.25, Cat. #: NT38-344, $260), and found that the latter produced images of better quality. The condenser used was a Zeiss system, N.A =1.4.

**D. Camera**

We tested two inexpensive, black and white, board-level CCD cameras: the 20VC3617 model by Videology ($20, available from BG Micro), and the STC-130CS model by Sentech ($94, available from Aegis, CA). Although it was possible to visualize microtubules using the Videology camera, the best images were obtained with the Sentech model, making it our choice throughout. This camera has a 1/3″ CCD with 510×492 active pixels, measures 35mm×35mm, and is easy to mount. The video output was connected directly to a frame grabber (PCI-1405, National Instruments, $679) installed in a tower PC.

**E. Image acquisition and processing**

Video images were acquired using custom-written software using LabView, Version 7.1.1, equipped with the add-on package Vision (National Instruments). This enabled us to implement two essential, real-time image-processing methods for MT visualization: background subtraction and frame averaging. Using the background-free images produced with the LabView code (available upon request), it was possible to align the microscope until microtubules were clearly distinguished, at which point we collected images at a typical rate of ~1 frame per second. Data was stored as an AVI movie file. To enhance image contrast and visibility further, subsequent video processing was performed, using the freeware programs VirtualDub 1.7.3 (http://www.virtualdub.org/) and ImageJ (http://rsbweb.nih.gov/ij/). VirtualDub was mainly used to crop and resize large video files, whereas ImageJ was used to implement image extraction operations that transformed the gray-scale images to binary masks, allowing automated tracking of microtubules upon displacement (see Section V).

**III. SAMPLE PREPARATION**

In this section we describe how to immobilize microtubules on the surface of a microscope coverslip, the *in-vitro* assay used in most of our experiments. All preparations and experiments are performed at room temperature, except when noted. Although microtubules are dynamic polymers in the cellular environment (undergoing periods of growth followed by shrinking events)[3], we work with stable microtubules, which can be produced by means of the anti-cancer drug Taxol. In the Appendix we provide a protocol to prepare reconstituted, stable microtubules from a commercial source. We strongly recommend that students be advised by researchers in a biology or chemistry lab who are familiar with the required handling of reagents and equipment (microcentrifuge tubes to hold solutions, micropipettes to dispense samples, etc.).

To prepare a sample of microtubules for the microscope, a flow cell is made by placing two pieces of double-sticky tape on a 3″×1″ microscope slide, creating a channel ~5 mm wide, and covering it with a coverslip (see Figure 2 in Ref. 13). If the coverslip is used directly from the box without any surface pretreatment, attachment of microtubules to the surface, although present, is minimal. Instead, we suggest treating the coverslip with a plasma cleaner (Harrick Plasma) for 5 min at ~1 Torr. Once the flow cell is made, about ~20 µl of microtubules in solution (see Appendix) are dispensed on one side of the channel using a micropipette, while filter paper (or a pipette tip connected to a vacuum line) suctions the liquid through the channel. We recommend adding a dilute sample of polystyrene beads (300-800-nm diameter, Spherotech) to the microtubule solution, prior to flowing into the chamber (a final bead dilution of 1/10,000 is sufficient). As some beads stick to the surface and cover it sparsely, this helps significantly in finding the coverslip surface during alignment of the microscope. After the MT solution is introduced into the chamber, the sample is left to incubate for 5 min, allowing for MTs to bind to the glass via non-specific adsorption. The flow cell is then washed, by injecting ~40 µl of PEMTAX buffer solution (see Appendix) through the channel, thus removing any unbound MTs or beads. Finally, the openings of the flow cell are sealed using vacuum grease or nail polish.

## IV. ALIGNMENT OF THE MICROSCOPE

Proper alignment of the microscope optimizes the capabilities of the instrument, and is essential for the success of the experiments described in this paper. In bright-field microscopy, optimal image resolution, contrast and fidelity are achieved by means of *Kohler illumination*. An extended discussion on the principles of Kohler illumination, together with a detailed procedure of alignment, can be found elsewhere[8]. Briefly, the microscope lenses are adjusted, starting with the light source, to satisfy three conditions. First, the collector lens images the light source onto the rear focal plane of the condenser, where the condenser diaphragm is located, creating a set of plane waves illuminating the sample. Second, the condenser lens images the field diaphragm (located near the collector lens) in the sample plane, setting the light source plane reciprocal to the sample plane. This defocuses non-uniformities of the extended light source, and therefore maximizes homogeneous illumination across the sample. Finally, the objective lens and subsequent optics image the sample onto the camera (Figure 1a).

In Kohler illumination, two sets of conjugate planes are formed. The first set includes the field diaphragm, the sample plane and the camera plane. The second set is formed by the light source position, the rear focal plane of the condenser and the back focal plane of the objective. Furthermore, these two sets are reciprocal, thus allowing the imaging of features on the specimen, not those present on the light source. Finally, this type of illumination provides the additional advantage that the working numerical aperture of the condenser can be adjusted independently, enabling use of the light source as either extended or effectively point-like. This allows adjustment of contrast, necessary for imaging fine specimen features.

To align the microscope, we use an iterative procedure. We begin by adjusting the collector lens (inside the 2″ lens tube) such that an image of the LED array can be seen

~1-2 m away from the lens, nearly collimating the light source. The illumination assembly is then mounted on top of the microscope. An alignment flow cell is prepared by incubating for 5 min a ~1000-fold dilution of beads in PEM buffer (Appendix), washing and sealing the cell, and mounting it on the microscope using immersion oil. Next, the condenser diaphragm is opened entirely, the condenser position lowered to provide enough illumination, and the objective adjusted so that clear images of surface-stuck beads can be seen with the camera. We then close the field diaphragm almost entirely, and the condenser is adjusted such that the diaphragm's edges are clearly visible on the field of view. At this point the distance between the collector lens and the position of the condenser diaphragm (*d*) is measured (to within ~1 cm). In the next iteration, we demount the illumination assembly and adjust the position of the collector lens (using the threaded rings that secure it within the 2″ lens tube), such that an image of the LED array can be seen at the measured distance *d* away from the collector lens. The illumination assembly is then mounted again, centered and secured. We repeat adjustments of the objective (to image surface-stuck beads), and of the condenser (to image the edges of the nearly-closed field diaphragm). Finally, the field iris is opened just so that it is no longer in the field of view. At this point, our microscope is aligned to Kohler illumination. One last step of alignment is of paramount importance for MT-visualization: we close the condenser iris almost entirely (~85% closed), yielding and effective point-like illumination source—enhancing contrast. At this stage we are ready to image sub-diffraction-sized biological polymers.

## V. IMAGING INDIVIDUAL MICROTUBULES

A new flow cell with MTs and only a few beads attached to the surface is prepared (following the procedure detailed in Section III), and mounted on the microscope. While viewing with the camera (or through the eyepiece), the coverglass surface is found with the help of stuck beads or dirt. No beads should be free in solution (they should have been removed during the washing step in the preparation of the flow cell), as they will interfere with the background subtraction procedure. After focusing the objective on stuck beads, the condenser and the field diaphragm are re-adjusted for Kohler illumination.

Focusing the microscope to stuck beads is enough to image the surface; however, the MTs cannot be seen without image processing (Figure 3a). Therefore, the sample chamber is moved with the XY translation stage to a region where no beads are present, the objective is adjusted to defocus off the surface, and a number of background images taken (typically 200) and averaged, providing a single background frame which is then subtracted from all subsequent frames. The second step to visualize MTs in real-time is to perform frame averaging. While MTs can barely be seen using only one background-free video frame, they are easily seen with ~10 or more averaged frames. As mentioned earlier, we capture video and implement these operations using custom-made LabView code. To complete alignment, we slowly refocus the objective back to the surface while looking carefully at the background-free, averaged images: individual microtubules should appear as elongated objects, roughly aligned in the direction of the flow cell channel (Figure 3b-d).

The procedure described above provides real-time video images of clearly visible microtubules. The software ImageJ allows further off-line image-processing, making it possible to automatically find and outline microtubules with particular orientations (Figure 4). Many image operations are available using ImageJ, and students are encouraged to experiment with them. Here, we describe only a few of the processes that improved significantly the quality of our images. First, we implemented spatial filtering by using the macro *Process→FFT→Band-pass filter*. The program computes the Fast Fourier Transform of the image, removes specified structures corresponding to certain spatial frequencies, and applies the inverse Fourier transform, thus returning the filtered image with much reduced noise or background. We typically suppress large structures down to 40 pixels, and small structures up to 5 pixels. To enhance features only in the vertical (horizontal) direction, the option *Suppress stripes Horizontal* (*Vertical*) was used. Second, we made extensive use of smoothing and enhancement by means of spatial convolution routines. In one dimension, the convolution operation for any two continuous functions $f(x)$, $g(x)$ is defined as: $f \otimes g(x) = \int f(x) \times g(x-y) dy$. The 2D, discrete operation used within the context of image processing is: $B_{i,j} = \sum_m \sum_n I_{i-m, j-n} M_{m,n}$, where $B_{m,n}$ is the processed image, $I_{i,j}$ is the original image, and $M_{m,n}$ is the convolution mask or kernel[8]. The point-by-point convolution operation modifies a target pixel by adding contributions from neighbor elements, contributions that are specified by the kernel, a 2D square matrix (typically 3×3 or 5×5). Kernels we used repeatedly to enhance our images acted preferentially on a particular direction; for example:

$$\begin{pmatrix} 0 & 1 & 0 \\ 0 & 1 & 0 \\ 0 & 1 & 0 \end{pmatrix} \quad \text{or} \quad \begin{pmatrix} 1 & 0 & -1 \\ 2 & 1 & -2 \\ 1 & 0 & -1 \end{pmatrix}.$$

The first example corresponds to smoothing (along the vertical direction), and can be readily implemented through *Process→Filters→Convolve*. The second example enhances image (vertical) features or boundaries, creating the effect of shadowing. The command *Process→Shadow* implements this last convolution filtering. When used with our bright-field microscope images, the *FFT-band-pass filter*, *Convolve* and *Shadow* macros create impressive images of single microtubules whose quality is strikingly similar to those obtained by high-end methods, such as Nomarski-DIC (Figure 4d).

After image enhancement, the signal-to-noise ratio of microtubule images is large enough to apply a threshold (*Image→Adjust→Threshold*) and retain only the most salient features (Figure 4e). The threshold-adjusted images are converted to binary (*Process→Binary→Make binary*) and subsequently into mask frames (*Process→Binary→Convert to Mask*), thus allowing automated feature extraction (Figure 4f). This last operation can be performed with the command *Analyze→Analyze Particles*, which scans for and retains features with a minimum size specified in px$^2$ (we typically use ~500 px$^2$), and circularity (between 0-1). Enabling the option *Show Mask* displays the extracted features in a new window.

## VI. A KINESIN MOTILITY ASSAY

Kinesin is an elongated protein that ferries cargo inside cells, using microtubules as freeways[3]. To achieve this feat, the molecule is powered by two tiny (~4 nm) identical motor subunits (called 'heads') that transform the energy of ATP hydrolysis into motion. With striking similitude to a person walking, kinesin uses its two motor heads as feet: binding to and unbinding from microtubules and thus stepping along, always towards a preferential end of the microtubule (called the plus-end). We show how it is possible to visualize and track kinesin-powered motion using our instrument, by fixing kinesin motors upside down in a coverslip and allowing the heads to bind and translocate free microtubules in the presence of ATP (the so-called 'gliding-filament' geometry, Figure 5a).

A flow cell is prepared, and a sample of ~20 µl of reconstituted kinesin solution (see Appendix) is introduced into the channel and incubated for ~10 min at room temperature. Then, the flow cell is washed with ~200 µl of MOTILITY buffer to remove any unbound motors, and a solution of microtubules diluted in MOTILITY buffer is introduced in the cell (see Appendix). Finally, the chamber is sealed and mounted on the microscope, after which the procedure to image microtubules is the same as before. Once microtubules are clearly visible using background-free, averaged images (we averaged 30 frames), their motions should be noticeable after a few seconds. It is impressive to watch microtubules undergoing constant displacements, over distances corresponding to tens of microns, and occasionally changing directions and even circling around sticky points. It is important to notice that we use high kinesin concentrations, such that multiple motors are involved in displacing a single microtubule.

We typically collect 50-150 frames of moving MTs and save them into an AVI file. Using the off-line, digital-image processing detailed in Section V, we extracted the shapes corresponding to individual kinesin-driven microtubules in all frames of the movie file (Figure 5b-d). The center of mass coordinates of moving microtubules were obtained with ImageJ, by setting *Analyze→Set Measurements→Center of Mass* and running the *Analyze→Analyze Particles* macro, with the *Display Results* option enabled. The automated extraction procedure yields center of mass coordinates (in pixels) that need to be converted into distance units. This was done by imaging a micrometer-scale slide with the microscope and counting the number of pixels contained between scale markers, from where the pixels to microns conversion factor was established.

A typical MT trajectory is shown in Figure 5d, where a single MT is outlined, and the corresponding distance *vs.* time graph (Figure 5e) shows variations in the translocation speed. Typical MT velocities observed (~200 nm/s) are ~2-fold lower than expected for this particular kinesin protein[18], which we attribute to the use of non-specific binding of kinesins to the coverslip. Tracking of MTs could be extended for over ~150 sec, limited mainly by drift in the microscope that defocused imaging.

## VII. CONCLUSIONS AND FINAL REMARKS

Using a simple setup, we have demonstrated that individual microtubules can be visualized through video-enhanced, bright-field microscopy. The experiments described here should contribute to physics and biophysics courses appropriate for advanced undergraduate and graduate students in physics and biology programs. To further extend and improve upon this work, the motility assay could be taken to the single-molecule regime, where the coverslip surface is sparsely coated with motors, such that only a single kinesin molecule moves a microtubule. This would allow a measurement of the flexibility of kinesin by scoring the angular motions of microtubules[19]. Our setup could also be combined with optical tweezers optics and a bead assay[13], making it possible to measure the force developed by kinesin as it walks along the microtubule.

## APPENDIX: BUFFERS, MICROTUBULE AND KINESIN PREPARATIONS

The tubulin and kinesin used in our study were acquired from Cytoskeleton, provided as vials with lyophilized samples (tubulin, Cat. #: T238; kinesin, Cat. #: KR01). Although most of the necessary chemicals are available from Sigma, a viable option would be purchasing ready-made buffers, available from Cytoskeleton. In fact, Cytoskeleton offers pre-formed microtubules (Cat. # MT001) and a full kinesin motility kit (Cat. #: BK027). Below, we provide protocols that indicate how to re-suspend lyophilized tubulin and kinesin, how to polymerize microtubules, and which dilutions are needed for optimal viewing in the microscope. Whenever equivalent buffers or solutions are available from Cytoskeleton, we indicate the corresponding catalog part number in square brackets.

Buffers and solutions needed:

PEM buffer [BST01]: 80 mM Pipes, pH 6.9, 1 mM EGTA, 4 mM $MgCl_2$
TAXOL (Paclitaxel) [TXD01]: 2 mM in DMSO
PEMTAX: dilute 5 μl TAXOL in 1 ml PEM
GTP [BST06-001]: 100 mM in PEM
PEMGTP: dilute 1 μl GTP in 99 μl PEM
STORAGE buffer: 80 mM Pipes, pH 6.9, 50 mM potassium acetate, 4 mM $MgCl_2$, 2 mM DTT, 1 mM EGTA, and 20 μM ATP
MOTILITY buffer: 80 mM Pipes, pH 6.9, 50 mM potassium acetate, 4 mM $MgCl_2$, 2 mM DTT, 1 mM EGTA, 10 μM taxol, 2 mg/ml BSA, and 2 mM ATP

Microtubule polymerization:

Resuspend lyophilized tubulin by adding 100 μl PEMGTP, yielding 10 mg/ml tubulin (TUB). Then, mix: 60.8 μl PEMGTP + 2.2 μl DMSO + 4.8 μl TUB, and incubate the admixture at 37°C for 30 min to grow microtubules. In the meantime, mix 83.6 μl PEM + 1.0 μl GTP + 9.4 μl 65 g/l $NaN_3$ + 6.0 μl TAXOL (STAB). After the incubation time, stabilize microtubules by adding 8 μl STAB to polymerized tubulin. Stable microtubules can be kept at room temperature over several weeks. To view under the microscope, MTs are diluted in PEMTAX buffer, typically 1:100.

Kinesin-microtubule assay:

Resuspend lyophilized kinesin by adding 10 μl of STORAGE buffer, yielding 2.5 mg/ml kinesin. For long-term storage (~6 months) we recommend adding 10 μl glycerol, mixing, and storing at −20°C. To use in the motility experiment, 1 μl of resuspended kinesin is diluted into 99 μl MOTILITY buffer. Although kinesin is a robust protein, it can only survive at room temperature for a few hours, and we therefore suggest keeping kinesin samples in ice until use, and prepare new samples after ~1 hr of observation. For this experiment, MTs are diluted in MOTILITY buffer, typically 1:200.

**ACKNOLWEDGEMENTS**

This work was supported by a grant from the National Institutes of Health (to S.M.B.). We thank members of the Block Lab for helpful discussions.

[a] Present address: Instituto Potosino de Investigación Científica y Tecnológica. Camino a la Presa San José 2055, CP 78216. San Luis Potosi S.L.P. Mexico
[b] Electronic mail: bgutierrez@ipicyt.edu.mx
[c] Electronic mail: sblock@stanford.edu

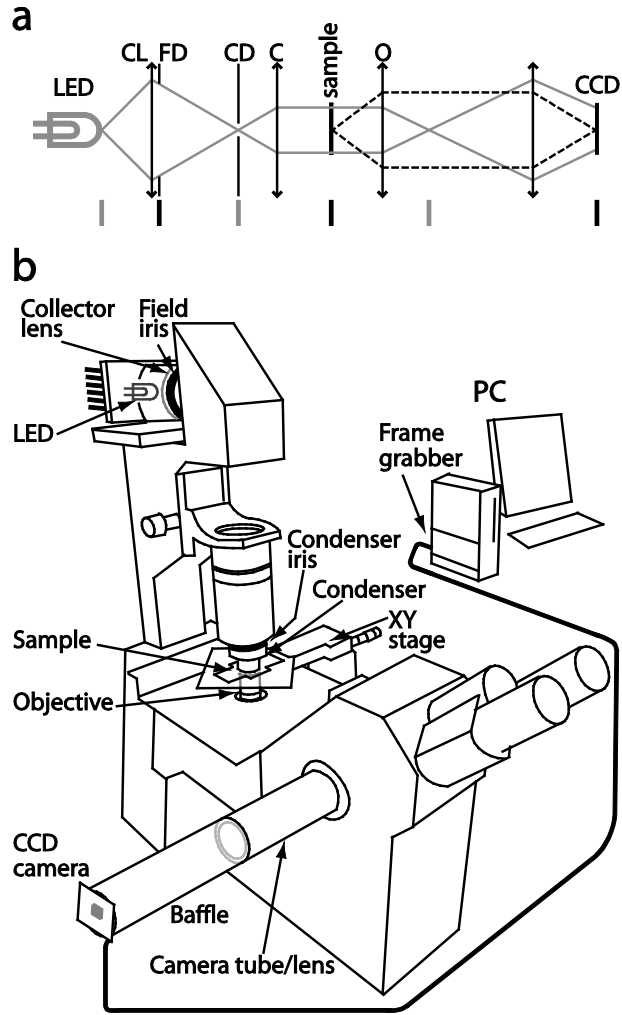

**Figure1.** Schematic of the bright-field microscope used for microtubule visualization. (a) Arrangement of the various microscope lenses (lines with end arrows) and apertures (vertical segmented lines) to Kohler illumination (see main text). The grey lines correspond to the illumination light, whereas the black dotted lines represent light scattered by the sample and imaged by the objective. Vertical lines of the same color (grey or black) indicate a set of conjugate planes. CL: Collector lens, FD: Field diaphragm, CD: Condenser diaphragm, C: Condenser, O: Objective. (b) Microscope components and the video-computer interface. The LED source and the collector lens were mounted on a lens tube housing, which was bolted to a heat sink. The camera tube with the imaging lens was adjusted slightly to obtain the desired magnification. A baffle covered the path between the lens tube and the CCD camera, blocking any room light.

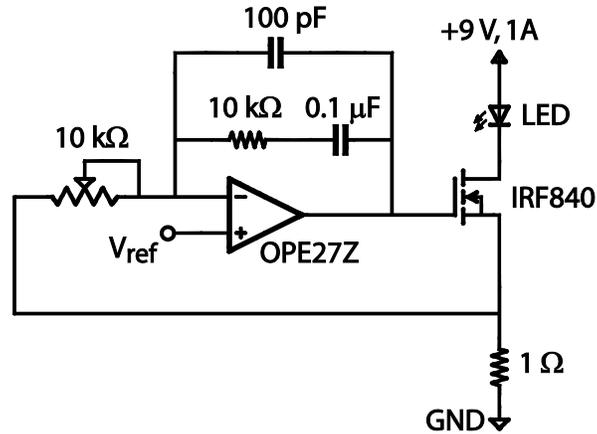

**Figure 2.** Circuit diagram of the LED current controller. Current through the LED is controlled by a high power MOSFET, and monitored by a 1Ω, 1W resistor. The monitor voltage is fed back to a low-noise operational amplifier (in a Proportional-plus-Integral, negative-feedback mode), whose output controls the MOSFET gate, completing the servo loop. A reference voltage sets the desired current, while the potentiometer allows adjustment of feedback gain, preventing oscillations.

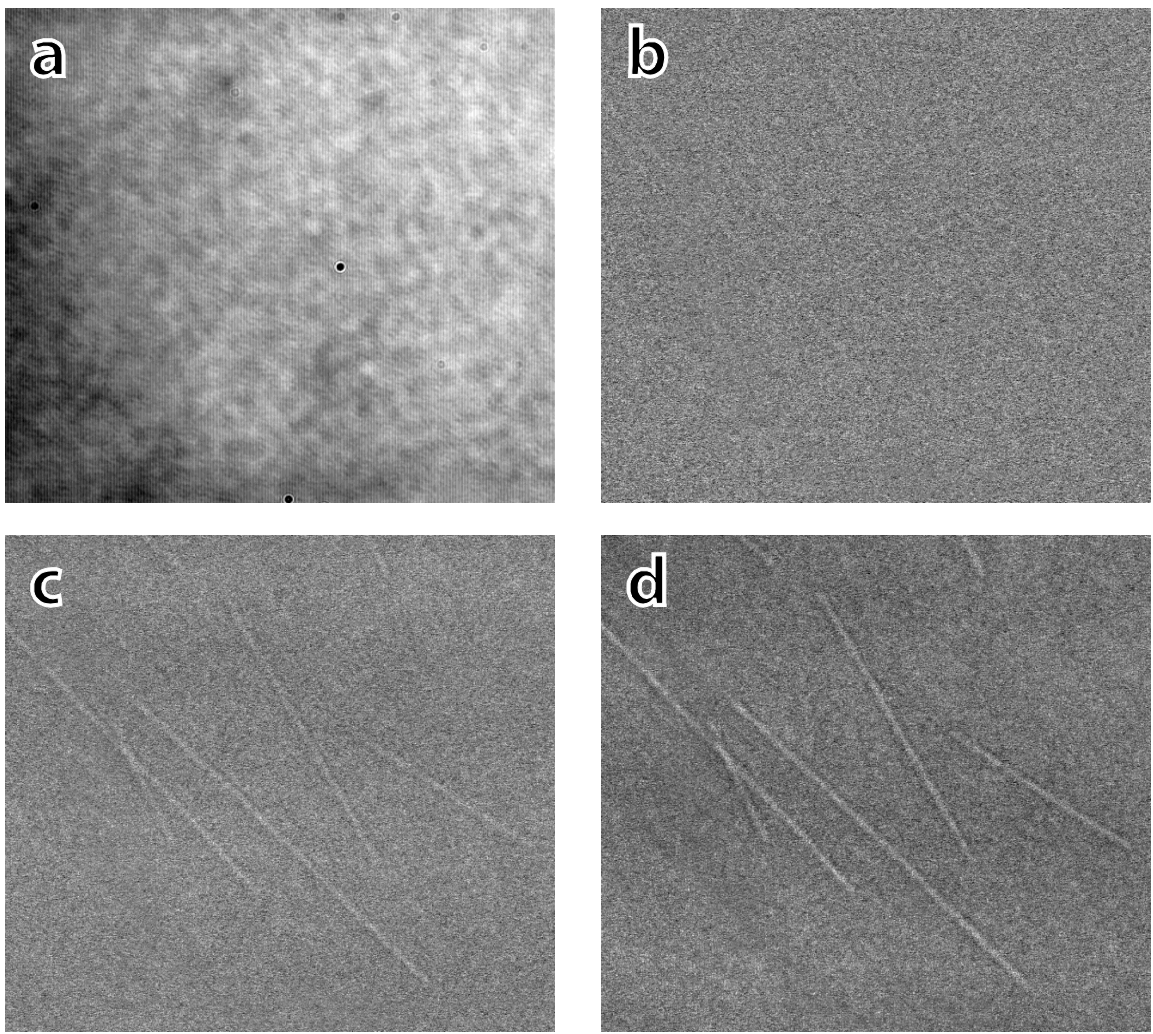

**Figure 3.** Real-time video images of microtubules bound to a coverslip. (a) Raw CCD image. (b) Single background-free frame. (c) Average of 10 background-free frames. (d) Average of 30 background-free frames. Individual microtubules are clearly visible after averaging ≥10 frames (corresponding to ~3 frames per second). Field of view: 20×20 μm.

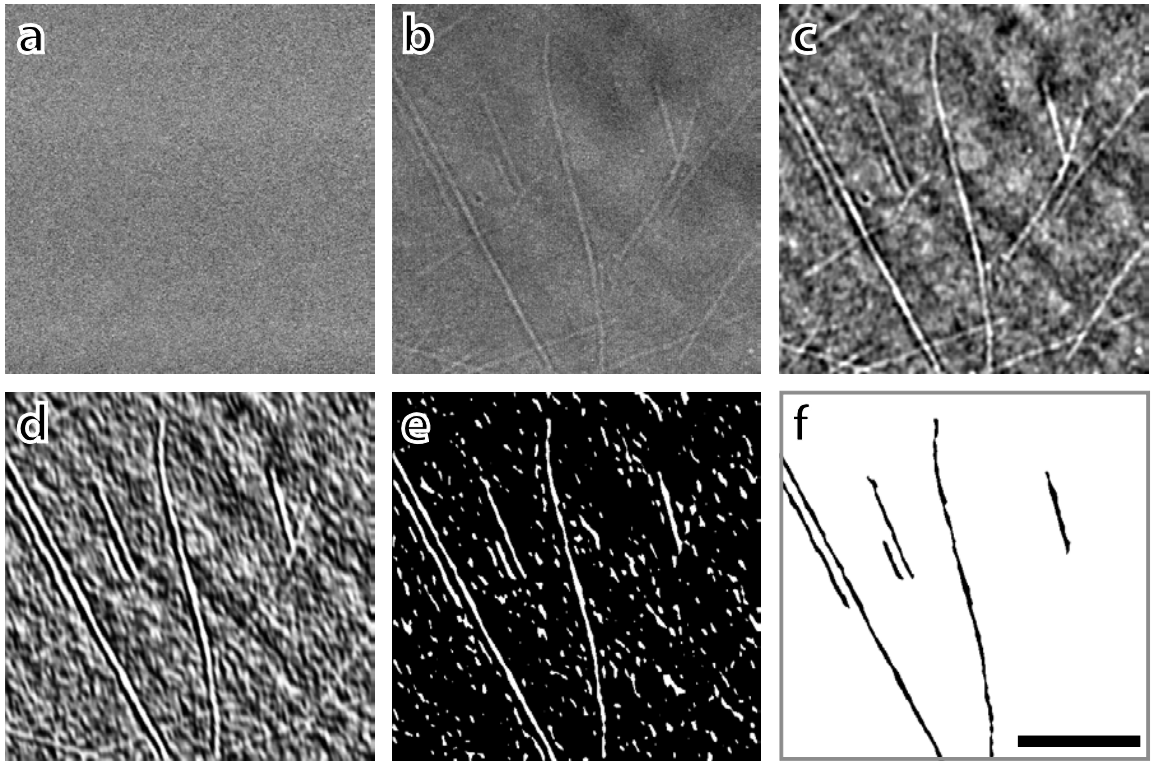

**Figure 4.** Digital processing of video images with ImageJ. (a) Single background-free frame. (b) Reference image used for sequential modifications: Average of 30 background-free frames. Individual microtubules can be observed in various orientations. (c) Using FFT band-pass filtering, large structures down to 40 pixels, small structures up to 5 pixels, and horizontal features are largely suppressed. (d) Convolution-filtered image, using 'smoothing' and 'shadowing' kernels in a diagonal direction. Note that horizontal structures are almost inexistent. (e) Application of a threshold operation retains only high-intensity features, from where a binary mask is obtained. (f) Extraction of large features from (e) eliminates all noise and retains shapes of microtubules with a particular orientation. See text for details. Horizontal scale bar: 5 μm.

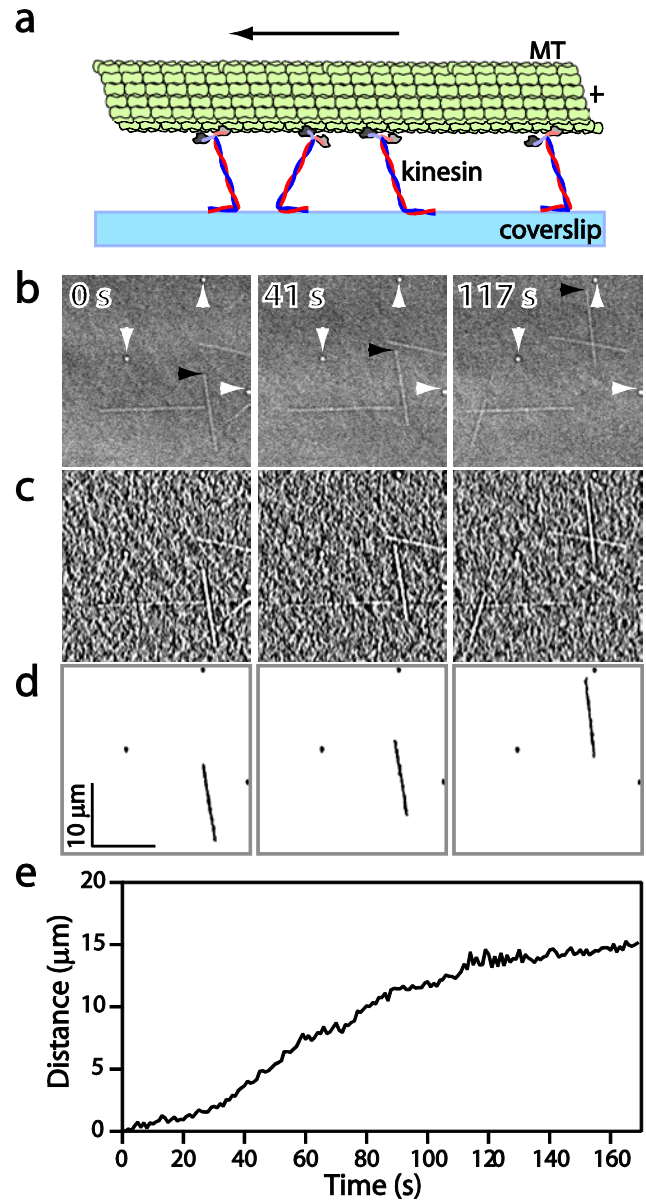

**Figure 5.** Measuring kinesin motility in a gliding-filament assay. (a) Cartoon of the gliding-filament assay: kinesin molecules fixed to the coverslip bind to a microtubule by the heads in the presence of ATP. As kinesin molecules start 'walking' towards the microtubule 'plus end' (+), the microtubule is displaced in the opposite direction (horizontal arrow). (b) Representative image sequence showing a single microtubule (black arrowhead) as it displaces over the surface driven by kinesin molecules. Three small particles fixed to the surface (white arrowheads) serve as stationary fiducial marks. Each image corresponds to the average of 30 background-free frames. (c) Images shown in (b) after FFT band-pass filtering, and convolution filtering. (d) After feature extraction, the shape of the moving microtubule is clearly seen, along with the three reference particles. (e) Distance traveled by the microtubule as a function of time, obtained from centroid tracking after feature extraction. Frame rate used in this experiment: 1 fps.